\documentclass[journal,twoside,web]{ieeecolor}
\usepackage{generic}
\usepackage{cite}
\usepackage{amsmath,amssymb,amsfonts}
\usepackage{algorithmic}
\usepackage{graphicx}
\usepackage{algorithm,algorithmic}
\usepackage{hyperref}
\hypersetup{hidelinks=true}
\usepackage{textcomp}
\usepackage{multirow}
\usepackage{makecell}
\newtheorem{lemma}{Lemma}
\usepackage{tikz}
\newcommand\submittedtext{%
  \footnotesize This work has been submitted to the IEEE for possible publication. Copyright may be transferred without notice, after which this version may no longer be accessible.}

\newcommand\submittednotice{%
\begin{tikzpicture}[remember picture,overlay]
\node[anchor=south,yshift=10pt] at (current page.south) {\fbox{\parbox{\dimexpr0.65\textwidth-\fboxsep-\fboxrule\relax}{\submittedtext}}};
\end{tikzpicture}%
}

\newcommand{\rebuttal}[1]{{\leavevmode\color{black}#1}} 

\def\BibTeX{{\rm B\kern-.05em{\sc i\kern-.025em b}\kern-.08em
    T\kern-.1667em\lower.7ex\hbox{E}\kern-.125emX}}
    
\usepackage{array, booktabs}
\newcolumntype{B}{!{\vrule width 1.4pt}} 
\newcommand{\thickhline}{\noalign{\hrule height 1.4pt}}

\begin{document}
\title{Fast and Robust Stationary Crowd Counting \\with Commodity WiFi}
\author{Mert Torun, \IEEEmembership{Member, IEEE}, Alireza Parsay, \IEEEmembership{Member, IEEE}, and Yasamin Mostofi, \IEEEmembership{Fellow, IEEE} \vspace{-18pt}
\thanks{This work was supported in part by NSF CNS award 2226255, and in part by ONR award N00014-23-1-2715.}
\thanks{Mert Torun, Alireza Parsay, and Yasamin Mostofi are with the Department of Electrical and Computer Engineering, University of California, Santa Barbara (e-mail: merttorun@ucsb.edu; alirezaparsay@ucsb.edu; ymostofi@ece.ucsb.edu).}
\thanks{Approval of all ethical and experimental procedures and protocols was granted by the Institutional Review Board (IRB) Committee at UCSB.
}
}

\maketitle
\submittednotice

\begin{abstract}
This paper introduces a novel method for estimating the size of seated crowds with commodity WiFi signals, by leveraging natural body fidgeting behaviors as a passive sensing cue. Departing from prior binary fidget representations, our approach leverages the bandwidth of the received signal as a finer-grained and robust indicator of crowd counts. More specifically, we propose a mathematical model that relates the probability density function (PDF) of the signal bandwidth to the crowd size, using a principled derivation based on the PDF of an individual’s fidget-induced bandwidth. To characterize the individual fidgeting PDF, we use publicly available online videos, each of a seated individual, from which we extract body motion profiles using vision techniques, followed by a speed-to-bandwidth conversion inspired by Carson’s Rule from analog FM radio design.
Finally, to enhance robustness in real-world deployments where unrelated motions may occur nearby, we further introduce an anomaly detection module that filters out non-fidget movements. We validate our system through 42 experiments across two indoor environments with crowd sizes up to and including 13 people, achieving a mean absolute error of 1.04 and a normalized mean square error of 0.15, with an average convergence time of 51 seconds, significantly reducing the convergence time as compared to the state of the art. Additional simulation results demonstrate scalability to larger crowd sizes. Overall, our results showcase that the proposed pipeline enables fast, robust, and highly accurate counting of seated crowds.
\end{abstract}


\section{Introduction} \label{sec:Introduction}

\IEEEPARstart{P}{assive} human sensing and activity-aware smart environments have emerged as prominent areas of research in recent years. The widespread availability of sensing technologies (such as computer vision, radio frequency (RF), and acoustic) in everyday environments has enabled new opportunities for the ubiquitous monitoring of human presence and activities. The common goal of passive human sensing research is to enhance safety, healthcare, and quality of life without requiring direct interaction from the users. \rebuttal{ Researchers have utilized these modalities for a variety of applications, such as fall detection \cite{lee2021deep,ji2022sifall,abro2024multi}, early diagnosis of medical conditions and smart healthcare \cite{parsay2025gait, liu2022monitoring, alazeb2024effective,parsay2025comparative}, activity recognition \cite{singh2021deeply, adaimi2021ok, webber2021human}, person identification \cite{korany2020multiple, zhang2022wi}, and occupancy estimation \cite{korany2021counting, choi2022wi,xu2021crowd,pallaprolu2024crowd}, among others.}

Occupancy estimation and crowd counting, in particular, have received considerable attention from the research community. Estimating the number of occupants is essential for energy efficiency, resource allocation, and safety in indoor environments such as offices, classrooms, and healthcare facilities. Accurate occupancy information enables intelligent control of heating, ventilation, and air conditioning (HVAC) systems and supports rapid decision-making during emergencies. It also facilitates public health-conscious strategies, such as occupancy limits or social distancing.

To enable occupancy analytics, different sensing modalities have been used in recent years. Vision-based methods, for instance, have been utilized heavily for occupancy estimation tasks, particularly through the use of machine learning techniques. Common approaches include using convolutional neural networks (CNNs) \cite{zhang2016single} or density maps \cite{liu2018decidenet} to perform crowd counting. To further enhance coverage and robustness, other studies have explored multi-camera systems that address occlusions and increase spatial visibility \cite{zhang2021cross, zhang20203d}. Despite their success in crowd counting performance, vision-based systems face significant limitations, including privacy concerns, dependence on line-of-sight, and the need for consistent lighting, highlighting the need for alternative crowd sensing modalities.  

More recently, researchers have used radio frequency-based solutions for occupancy estimation and crowd counting to address the limitations of vision-based methods.  For instance, \cite{hsu2023novel} proposed a mmWave-based feature extraction method for crowd counting, using a k-nearest-neighbor classifier. Similarly, \cite{hu2024mmcount} introduced mmCount, a stationary crowd-counting system based on 2D heatmaps generated with mmWave radar. Sakhnini et al. \cite{sakhnini2024crowd} proposed a mmWave-based metric correlated with large crowd sizes. Deep learning \cite{choi2021deep} and transformer-based \cite{choi2024radar} approaches have also gained attention for radar-based crowd counting. In general, although radar-based methods are suitable for specific applications, they often require either customized hardware or more expensive commodity devices as compared to WiFi. Moreover, they can experience severe blockage as compared to WiFi, limiting their practicality as a ubiquitous solution.

In contrast, WiFi-based sensing has gained traction in the research community due to the widespread availability of WiFi infrastructure in everyday environments and its through-wall sensing capabilities. Earlier attempts in crowd counting ~\cite{depatla2018crowd,depatla2015occupancy} proposed a mathematical framework for counting moving crowds. More recent approaches \cite{di2016trained, zhang2020wicrowd, zong2020device, zhou2020device, wang2021crowd, zou2017freecount} have also used machine learning for this purpose, albeit with limited generalizability, as collecting diverse crowd training data is challenging. Overall, the aforementioned WiFi-based crowd counting work are focused on counting moving crowds, and the methods are not applicable to the problem of interest in this paper, which is focused on counting a seated crowd.  

In this paper, we are interested in counting seated crowds with WiFi signals, a relatively underexplored problem in the RF sensing literature, as it is considerably more challenging due to the lack of major body motion. \rebuttal{In \cite{zhao2019deepcount,liu2017wicount, khan2022crosscount,guo2025rssi}, neural network pipelines are trained to count a number of seated or mobile people. These work, however, can suffer from limited generalizability, as collecting diverse crowd RF training data is challenging. Moreover, most of the aforementioned work (except \cite{guo2025rssi}) involve very small group sizes (e.g., up to five individuals) and are thus better categorized as multi-person counting rather than true crowd size estimation.} In recent years, stationary crowd counting with WiFi CSI has also been tackled using fidget duty cycles, incorporating queuing theory to model fidgeting statistics. Specifically, in \cite{korany2021counting}, we proposed this approach to estimate the size of a seated crowd, while \cite{jiang2023pa} adopts a similar methodology in the context of passenger counting. Using fidget-induced signal variations is a promising first step toward solving the challenging problem of counting seated crowds, where large body motions are absent. However, the aforementioned existing method typically requires long observation windows to accumulate sufficient fidgeting statistics, relying on the observation of several fidget and silent periods to form a stable estimate. Therefore, these binary duty cycle-based approaches, where signal variations are treated as on-off fidget activity indicators, lead to coarse-grained modeling that prolongs estimation convergence. 

In this paper, we take a different perspective on counting a seated crowd. More specifically, we propose a finer-grained approach that efficiently models the full amplitude of fidget-induced signal fluctuations, as opposed to an on-off duty cycle-based approach. This approach allows our system to capture more information about the fidgeting behavior of crowds and converge with significantly less observation time. 

We next present the contributions of the paper in more detail.

\textbf{Statement of Contributions:}

1. In this paper, we propose a novel WiFi-based approach for estimating the size of stationary crowds by exploiting the fidgeting behavior of individuals. Rather than relying on a binary representation for the fidgets, we propose that the bandwidth of the received signal offers finer-grained information about cumulative crowd behaviors while being robustly measurable in practical scenarios. We then develop a mathematical model for the PDF of the received signal bandwidth in the presence of a seated crowd and relate it to the total number of people.  This derivation is a function of the bandwidth PDF of a single person’s fidgets, which we need to characterize. We then demonstrate how to extract the individual bandwidth PDF from online, public videos, each of a seated individual, using common tools in the area of vision and a speed-to-bandwidth conversion inspired by Carson’s Rule from analog FM radio design. Overall, this provides a principled mathematical framework for inferring the crowd size based on signal bandwidth observations. The proposed method achieves markedly faster convergence than prior state-of-the-art techniques, as we will show.

2. In order to improve the robustness in real scenarios when other people can pass by, inducing motions that are not from the seated crowd, we further propose an anomaly detection module that effectively filters non-fidget movements. This module has the capability to detect and filter out non-fidget motions, which enhances the estimation performance under realistic conditions. To the best of our knowledge, this is the first stationary crowd-counting work to account for such anomalies, which are very likely to occur in real-life deployments.

3. We conduct $42$ experiments in two different indoor environments for various seated crowd sizes of up to and including $13$ people. Across all the experiments, our system achieves a mean absolute error of $1.04$ and normalized mean square error of $0.15$, with an average convergence time of $51$ seconds, a considerably smaller value than the state of the art, as we shall see. In addition to extensive experimentation, we also provide empirical simulation test results to further showcase the performance of our proposed pipeline with much larger crowds. Overall, our results confirm that the proposed methodology delivers fast, robust, and highly accurate seated-crowd counting.

\begin{figure*}[t]
    \centering
    \includegraphics[width=0.90\textwidth]{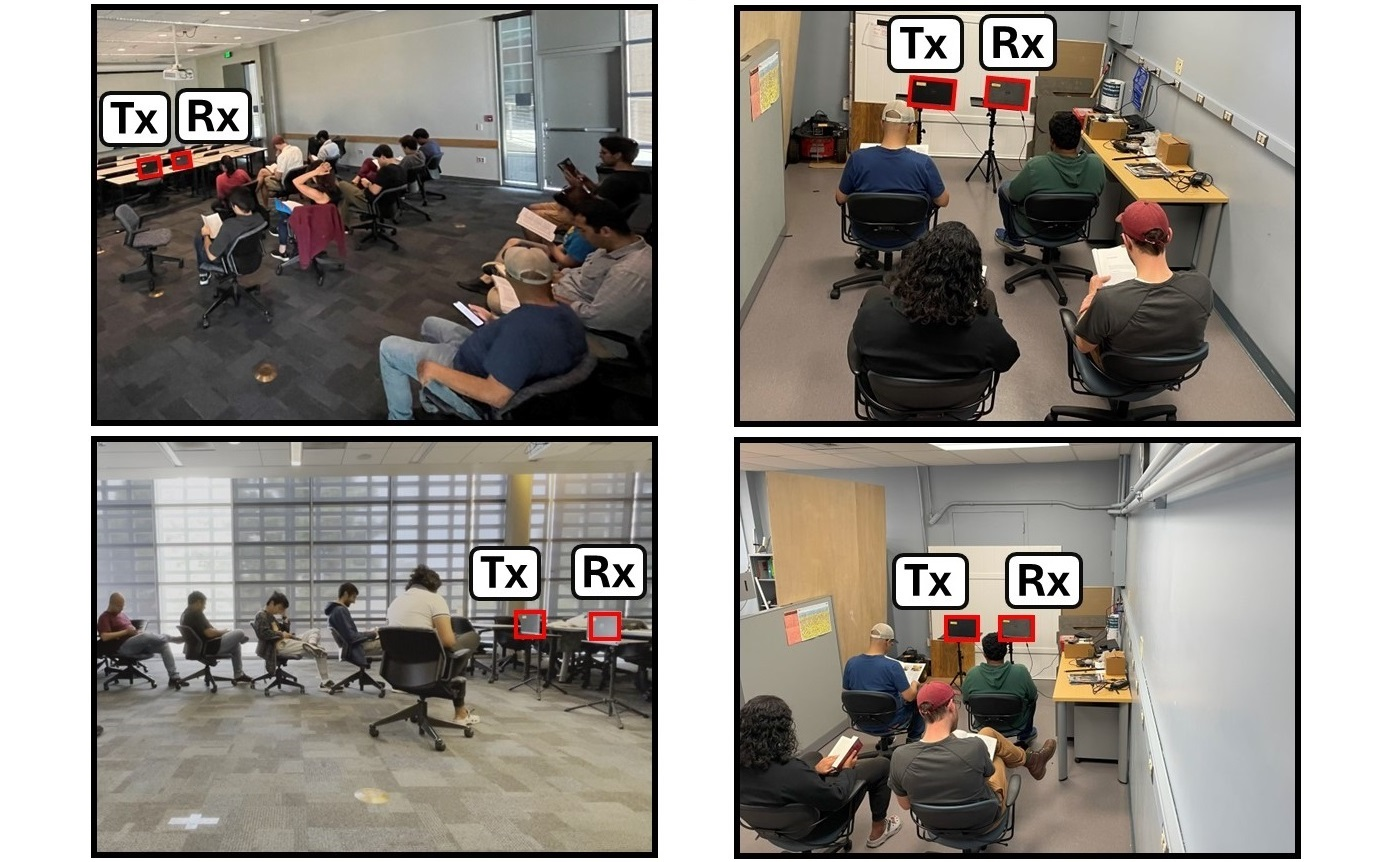} 
    \caption{Sample seated crowd scenarios (from our experiments) where WiFi transceivers of a pair of laptops are used for crowd counting: (left column) conference room, (right column) office environment.}
    \label{fig:1}
\end{figure*}


\section{Problem Formulation}
\label{sec:probformulation}

Consider an indoor environment where a seated crowd is engaged in an activity such as reading, as shown in Fig.~\ref{fig:1}. Even when seated and involved in a stationary task, individuals naturally exhibit subtle involuntary movements, such as adjusting their posture or repositioning limbs, referred to as ``fidgeting". These fidgets then cause variations in reflected WiFi signals, becoming observable through the Doppler effect in the received signals.  More formally, the complex baseband received signal can be expressed as follows~\cite{torun2023wiflex}:
\begin{equation}
\label{Baseband}
    s_b(t) = \alpha_\text{s}e^{j\theta_s} + \sum_{m \in \Omega} \alpha_m e^{j\left(\frac{2\pi}{\lambda}\psi_m \int v_{m}(t) dt + \frac{2\pi}{\lambda}d_m\right)},
\end{equation}
where $\alpha_\text{s}e^{j\theta_s}$ represents the combined effect of the direct path from the transmitter to the receiver and the reflections from static objects, $\alpha_m$ is the amplitude of the signal path reflected off the $m^{\text{th}}$ body part, $d_m$ is the path length at time $t=0$, $v_{m}(t)$ is the speed component of the $m^{\text{th}}$ body part at time $t$ along the perpendicular line to the ellipse whose foci are TX and RX, and $\psi_m=2\cos \phi_m$, where $\phi_m$ is the angle between the direction of $v_{m}(t)$ and the path from $m^{\text{th}}$ body part to the receiver or transmitter (see~\cite{korany2021nocturnal} for details on the geometrical definitions). The summation in Eq. \ref{Baseband} is taken over the body parts of an individual visible to the transceiver at time $t$, which we denote with set $\Omega$, and $\lambda$ is the wavelength. Commercial off-the-shelf devices, however, robustly measure only the received signal power or the phase difference. The following is then the approximation of the power of the received signal after DC removal~\cite{korany2021nocturnal}:
\vspace{-3pt}
\begin{equation}
\label{PowerApprox}
    p(t) =  \sum_{m \in \Omega} A_m\cos \left(\frac{2\pi\psi_m}{\lambda} \int v_{m}(t) dt + \Delta \mu_m \right),
\vspace{-3pt}
\end{equation}
where $A_m=2\alpha_s\alpha_m$ and $\Delta \mu_{m}=\mu_{m}-\mu_{d}$ is the difference between the initial phase of the path reflected off the $m^{\text{th}}$ body part and the direct path. It is noteworthy that a similar expression can be written for the phase difference~\cite{korany2021nocturnal}.

As can be seen from Eq.~\ref{PowerApprox}, the received WiFi signal captures motion information from each moving body part of every individual in the environment. Specifically, each fidget leaves a measurable trace that can be extracted from the received signal. This insight forms the basis of our passive sensing approach, which aims to estimate the crowd size by analyzing the statistical properties of cumulative motion behavior. We next discuss the steps to achieve this goal in detail.

\begin{figure*}[t]
    \centering
    \includegraphics[width=\textwidth]{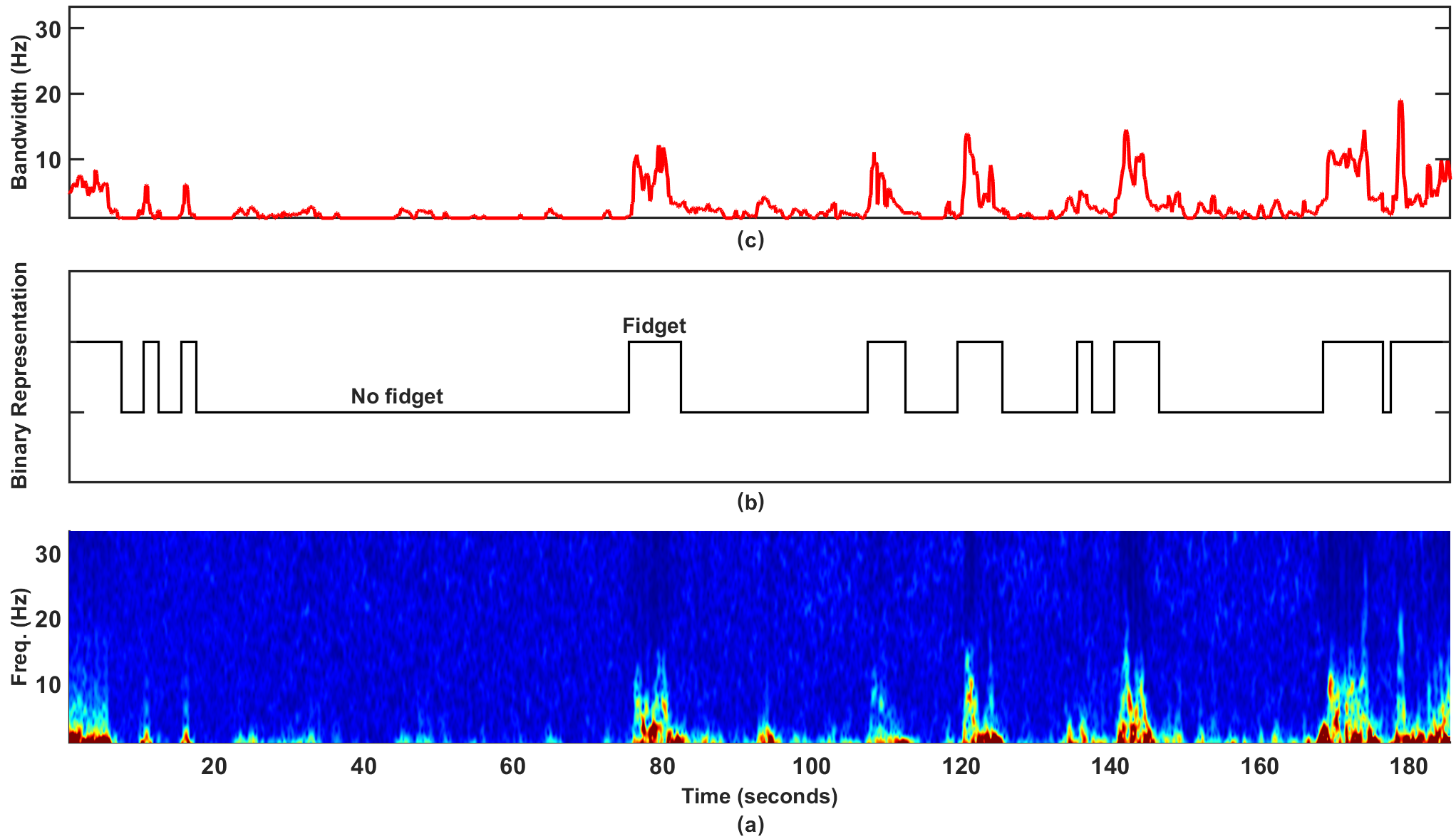} 
    \vspace{-8pt}
    \caption{Sample WiFi experiment from a stationary crowd \rebuttal{of three individuals} to demonstrate the difference between binary and non-binary bandwidth-based motion representations --  (a) spectrogram of the received WiFi signal capturing real-time fidget activities, (b) binary fidget representation obtained via thresholding, which provides a coarse, on/off encoding of motion, (c) proposed bandwidth-based representation enabling a more fine-grained and informative motion description, if it could be related to the crowd size in a principled way. In this paper, we show how to methodically extract the crowd size from the bandwidth.}
    \label{fig:1dto2d}
    \vspace{-8pt}
\end{figure*}

\section{A Richer Motion Representation} \label{sec:1dto2d}

The first step in designing a system capable of estimating stationary crowd sizes is to choose an effective representation of motion that is simple and robustly extractable, yet sufficiently expressive. As discussed in Section~\ref{sec:Introduction}, prior work in stationary crowd counting often models motion information using binary motion indicators that segment the received signal into discrete fidget and non-fidget periods, as can be seen in Fig.~\ref{fig:1dto2d} (b), forming what is known as a fidget duty cycle \cite{korany2021counting,jiang2023pa}. These methods infer the crowd size by analyzing the frequency and duration of such periods, operating on a coarse representation of motion. While conceptually straightforward and a great first step in solving this challenging problem, this binary encoding neglects a key aspect of the signal: the velocity of motion. In contrast, the bandwidth of the received WiFi signal captures information about the speed of underlying body part movements and can offer valuable insight for distinguishing between different levels of cumulative crowd motion, as we shall demonstrate. 

\textbf{Remark 1:} For brevity, in the rest of this paper, we refer to the state-of-the-art method for stationary crowd counting \cite{korany2021counting,jiang2023pa}, which leverages on-off fidget state detection, as the "binary" or "fidget duty-cycle-based" approach.

To illustrate the proposed underlying idea, consider Fig.~\ref{fig:1dto2d} (a), which shows a sample received WiFi signal spectrogram generated when a nearby stationary crowd was reading. Specifically, the figure captures the frequency vs. time information of the received signal, which is a function of different body parts' velocities. Fig.~\ref{fig:1dto2d} (b) then illustrates the motion representation with the binary approach, generated by thresholding in order to detect periods of activity. In \cite{korany2021counting}, we have then shown how to mathematically relate the dynamics of this binary representation to the crowd size, enabling crowd size inference. However, while this encodes when fidget motions occur, it collapses all amplitude variations, thereby erasing distinctions between weak and strong fidgets or overlapping motion events. This raises an important question: Is it possible to summarize fidget activities as a function of the total number of people while retaining meaningful information beyond binary representation?  Solving this problem enables the detection of overlapping fidgets, which in turn allows for significantly faster inference, as we shall see. 

A natural yet strong candidate to realize the envisioned velocity amplitude motion representation is the instantaneous received signal bandwidth (i.e., bandwidth as a function of time), which has been mathematically characterized in the RF sensing literature \cite{torun2023wiflex, korany2021nocturnal}, albeit for other purposes. More specifically, the bandwidth of the received signal, at a given time instant, is defined as the frequency below which the majority of the spectral energy is concentrated, as shown in Fig.~\ref{fig:1dto2d} (c). The bandwidth is directly influenced by body part speeds in the environment and is governed by the maximum of these contributing speeds (across a single person or multiple individuals). As a result, bandwidth serves as a compact, continuous, and velocity amplitude-aware feature that summarizes the most dominant motion activity present at any given time.

From a performance standpoint, this amplitude-aware representation retains fine-grained motion intensity over time. For instance, in binary encoding, multiple individuals fidgeting simultaneously appear indistinguishable from a single-person fidget, since both are reduced to the same active state. Our proposed approach, on the other hand, overcomes such an ambiguity by including the amplitude of the velocity in the representation in an efficient manner by utilizing the bandwidth. For instance, since bandwidth is governed by the maximum body part speed across all individuals, a high bandwidth observation can serve as a cue for a relatively larger crowd size. As a result, it has the potential to improve estimation efficiency. More specifically, binary methods require more alternating fidget and silent periods to accumulate sufficient statistical evidence (e.g., duty cycle ratios) before reaching their final count estimate. This requirement inherently introduces latency and increases the required observation time. In contrast, the continuous-valued bandwidth can capture the information of multiple people even when fidgeting simultaneously, which can result in a faster inference, as we shall see in Section~\ref{sec:results}. The main question, then, is how to relate instantaneous bandwidth measurements to the total number of people, in order to provide a base for crowd size inference. We next lay out the details of our proposed system, which shows how we can design a stationary crowd-counting pipeline that leverages bandwidth as its core motion representation.

\section{System Design} \label{sec:systemdesign}
In this section, we lay out the details of our proposed pipeline. More specifically, we develop a mathematical framework to estimate the size of a stationary crowd by modeling how human motion influences the bandwidth of the received RF signals. Specifically, we derive the probability distribution function (PDF) of the received signal bandwidth caused by crowd fidgeting, and relate it to the number of individuals. This derivation is a function of the bandwidth PDF of a single person’s fidgets, which we need to characterize. We then show how to extract the individual bandwidth PDF from online, freely-available videos of seated people, using a speed-to-bandwidth conversion inspired by Carson’s Rule from analog FM radio design. \textbf{This establishes a principled mathematical framework for estimating crowd size from signal bandwidth observations, eliminating the need for RF training data.}

We begin with a mathematical analysis of the received WiFi signal bandwidth, which characterizes how body part movements influence the signal’s frequency content. We then introduce a statistical model for cumulative crowd behavior, treating the received signal bandwidth as a random variable. This model allows us to methodically analyze how the bandwidth distribution changes as a function of the crowd size. Finally, we show how to get the bandwidth PDF of a seated individual using available online videos and a mathematical model introduced earlier for speed-to-bandwidth conversion.
\subsection{Relating Body Part Speeds of an Individual to Bandwidth} \label{bw}
In this part, we mathematically describe how the movements of the body parts of a single individual affect the received signal bandwidth. Consider Eq.~\ref{PowerApprox}, which describes the received power corresponding to the fidgeting movements of a single person, over a moving time window $T_{\text{mov}}$. The bandwidth of this signal as a function of time, corresponding to its Fourier transform over this window, can be characterized as follows:

\begin{lemma}
\label{lem:videoBW}
The bandwidth of the received signal of Eq.~\ref{PowerApprox} can be characterized as follows:
\begin{align}
\label{eq:videoBW}
BW(t) = \max_m \left( \frac{v_{\text{max},m}(t) \psi_m(t)}{\lambda} + f_{0,m}(t) \right),
\end{align}

where $f_{0,m}(t)$ is the bandwidth of the $m^{\text{th}}$ body part speed ($v_m(t)$) at time $t$, i.e., the bandwidth when directly taking the Fourier transform of $v_m(t)$ at time $t$, over $T_{\text{mov}}$, and evaluating its maximum spectral content.  Moreover, $v_{\text{max},m}$ is the maximum of $v_m(t)$ over the window of $T_{\text{mov}}$ centered at time $t$.
\end{lemma}

\textit{Proof:}
~The lemma can be proved using Carson's Rule as shown in~\cite{korany2021nocturnal,torun2023wiflex}.

Eq.~\ref{eq:videoBW} describes the behavior of the received signal bandwidth in the presence of multiple moving body parts and their corresponding velocity profiles. Beyond mathematically characterizing bandwidth dynamics, Eq.~\ref{eq:videoBW} also provides a practical method for estimating the received signal bandwidth, given access to the body part velocity profiles. Specifically, this observation enables estimating the bandwidth of the received signal from a given video of the person, using vision tools, as we shall see in Section~\ref{video-to-rf}. In the following subsection, we then present a mathematical model for the received signal bandwidth of a crowd of size $N$, building on the mathematical characterization of the bandwidth of an individual from Lemma~\ref{lem:videoBW}.

\subsection{Modeling the Observed Bandwidth of a Stationary Crowd of Size $N$} \label{bw-to-crowd}

Consider a crowd of size $N$. In this subsection, we model how the overall observed signal bandwidth evolves as a function of crowd size. We begin by treating the overall observed bandwidth as a random variable $BW$, and characterize the bandwidth behavior for an individual using the probability density function (PDF), denoted by $f_{BW}(x)$. Our goal is to then characterize the PDF of the observed bandwidth when $N$ individuals are present, which we denote by $f_{BW,N}(x)$.

We begin by expressing the cumulative distribution function (CDF) for a single individual as: $F_{BW}(x) = P(BW \leq x) = \int_{0}^{x} f_{BW}(\xi) \, d\xi$. In order to extend this model to a crowd of \( N \) individuals, we apply the following transformation. We assume that the bandwidth contributions from each individual are independent and identically distributed (i.i.d.), which is a reasonable assumption as individuals do not typically fidget in a correlated manner. Then, at each point in time, the bandwidth contribution from each person follows the distribution \( f_{BW}(x) \), with the observed final bandwidth for a crowd determined by the maximum among the $N$ independent contributions. This follows from the fact that signal components from different individuals combine linearly, and the dominant bandwidth is dictated by the fastest-moving body part across the group. Specifically, for i.i.d. random variables, we have the following for the cumulative bandwidth distribution for a crowd of \( N \) individuals:

\begin{align}
    \label{eq:cdfcrowd}
    F_{BW,N}(x) &= P\left( \max(BW_1, BW_2, \dots, BW_N) \leq x \right) \nonumber \\
    &= \prod_{i=1}^{N} P\left( BW_i \leq x \right) \quad \text{(} BW_i \text{ are independent)} \nonumber \\
    &= \left( P\left( BW \leq x \right) \right)^N \mkern1mu \text{(} BW_i \text{ are identically dist.)} \nonumber \\
    &= \left( F_{BW}(x) \right)^N,
\end{align}
where $BW_i$ is the bandwidth of the $i^{\text{th}}$ individual's signal, where we subsequently drop $i$ due to the i.i.d assumption.

Finally, in order to obtain the corresponding PDF for a crowd of $N$, we differentiate Eq.~\ref{eq:cdfcrowd}:
\begin{align}
    \label{eq:pdfcrowd}
    f_{BW, N}(x) &= \frac{d}{dx} F_{BW, N}(x) = \frac{d}{dx} \left( F_{BW}(x)^N \right) \nonumber \\
    &= N \cdot F_{BW}(x)^{N-1} \cdot f_{BW}(x)
\end{align}

Eq.~\ref{eq:pdfcrowd} methodically shows how the crowd size $N$ affects the PDF of the crowd bandwidth.  Moreover, this formulation also enables the computation of the expected bandwidth statistics for different crowd sizes, given the PDF of the bandwidth of a single person. During the inference stage, the observed bandwidth distribution can then be compared against the theoretical $f_{BW,N}(x)$, enabling the estimation of $N$ via statistical distribution matching.

In order to properly extract the crowd size from the theoretical distribution, $f_{BW,N}(x)$, we need to have a characterization for the PDF of the bandwidth of a single person $f_{BW}(x)$, which we show how to obtain in the next sub-section.

\begin{figure}[t]
    \centering
    \includegraphics[width=0.46\textwidth]{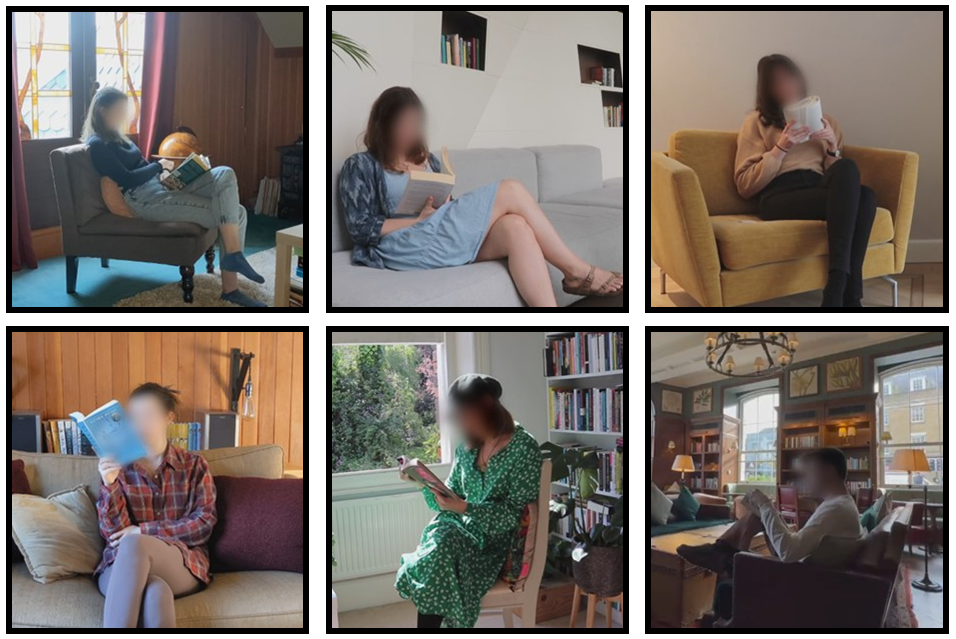} 
    \caption{Sample frames of videos of seated individuals from our compilation of such videos from public, online sources.  By choosing videos from various viewpoints while having a fixed TX-RX placement, our video-to-bandwidth pipeline effectively captures a broad range of relative angular configurations between the seated individual and the WiFi transceiver.}
    \label{fig:relativesittingorientations}
\end{figure}
\subsection{Harvesting Videos for RF Bandwidth Priors of an Individual} \label{video-to-rf}

As can be seen from Eq.~\ref{eq:pdfcrowd}, in order to estimate the crowd size $N$ from the PDF of the observed signal bandwidth, our method requires the PDF of the bandwidth when a single individual is fidgeting ($f_{BW}(x)$) as a prior. Once obtained, this enables a distribution-matching framework for identifying the most likely crowd size, using Eq.~\ref{eq:pdfcrowd}. However, manually collecting such WiFi bandwidth data of a single seated individual across multiple settings is a resource-intensive and prohibitive task. Moreover, such large-scale, task-specific WiFi datasets are not readily available. These challenges motivate us to turn to readily available vision datasets of seated individuals. Specifically, in this part, we propose a method for estimating the received WiFi signal bandwidth from a video of an individual, using Lemma~\ref{lem:videoBW}. This then enables us to estimate the PDF of the bandwidth of a seated individual using a video dataset, where each video features a seated individual.  We next provide more details on the method.

Recent advancements in computer vision have enabled the extraction of human body joint positions from standard videos. One such tool is MediaPipe Pose \cite{mediapipe}, developed by Google, which reliably estimates key joint positions from video frames in real-time, and is a widely used tool in the literature~\cite{kim2023human, parsay2025comparative, lugaresi2019mediapipe}. To estimate the received signal bandwidth that would have been observed if a WiFi transceiver were present in a video of a seated individual, we begin by extracting body joint trajectories using MediaPipe. We gather a dataset of $32$ unique videos featuring individuals engaged in reading, sourced from publicly available platforms (e.g., YouTube). Our dataset comprises $16$ hours and $51$ minutes of reading.

Frames of each video are fed to MediaPipe to extract the joint locations. For all visible joints, speed profiles (in pixels per frame) are computed by considering the differential over consecutive frames. To mitigate the effects of pose estimation noise, a low-pass filter is applied. The first challenge is converting these speed profiles from pixels per frame to meters per second for the depth at which the individual is located in the video. To address this, we propose using the average interpupillary distance of an adult as a scaling reference. More specifically, let $d_{ip}$ represent the estimated interpupillary distance of the individual in the video, measured in pixels. Using the average adult interpupillary distance of 63.36~mm \cite{dodgson2004variation}, we compute a pixel-to-meter conversion factor as $0.06336/d_{ip}$. The pixel-based joint speeds are then converted to meters per second using the derived scaling factor and the video’s frame rate, measured in frames per second (fps).

Having obtained joint velocity signals in physical units (m/s), we now examine how they can be used within the analytical framework of Eq.~\ref{eq:videoBW}. Several important considerations remain before Eq.~\ref{eq:videoBW} can be applied to the extracted speed profiles. The first requirement is to select an appropriate value for $\psi_m$, which depends on the system's configuration. In commonly-assumed configurations where the transmitter and receiver are positioned close to each other, relative to their distance from the individuals, the angle $\phi_m$ will be close to zero, and thus $2\cos \phi_m \approx 2$, which is what we shall use when considering a TX/RX configuration in each video.  

The next challenge is the realization of $v_m(t)$, which in real WiFi experiments represents only the component of motion perpendicular to the ellipse whose foci are the transmitter and the receiver. However, standard 2D videos provide access to only two spatial dimensions. Even if 3D videos were readily available, modeling each individual's orientation relative to the TX-RX link would remain a non-trivial task. To address this challenge in a practical way, we adopt an aggregation-based strategy by utilizing videos recorded from a variety of viewpoints. This approach introduces a natural variation in the projection of motion vectors onto the image plane depending on the camera viewpoint. For each video, we place the TX-RX configuration at a fixed far horizontal distance that yields $\psi \approx 2$. As illustrated in Fig.~\ref{fig:relativesittingorientations}, by incorporating videos from diverse viewpoints while having a fixed transceiver placement, our video dataset implicitly captures a broad distribution of relative sitting orientations with respect to the TX-RX pair. As discussed in Section~\ref{sec:probformulation}, the received WiFi signal bandwidth depends on the component of motion perpendicular to the TX-RX link (i.e., the projection of motion along this direction), therefore, by fixing the transceiver placement and having a diverse set of videos with various viewpoints we can generate a rich set of relative orientations without explicitly modeling them. This strategy further improves the generalizability of our pipeline by removing reliance on a fixed sitting orientation relative to the TX-RX link, which is especially important in real-world deployments where individuals may be seated in arbitrary orientations.

In summary, we apply Eq.~\ref{eq:videoBW} to each video and the associated speed profiles. The resulting $BW(t)$ estimates are aggregated into a pool to generate $f_{BW}(x)$. This bandwidth prior effectively captures a broad range of relative sitting orientations. In the next section, we describe how this video-to-bandwidth prior can be used to estimate the size of a seated crowd.

\subsection{Distribution Matching for Crowd Size Estimation}
Now that we have a method for generating a generalized and robust individual bandwidth distribution $f_{BW}(x)$, we can apply Eq.~\ref{eq:pdfcrowd} to compute the bandwidth PDF of a crowd of size $N$, $f_{BW,N}(x)$. Once these PDFs are constructed as a function of the crowd size, we can use them in the inference stage to estimate the true crowd size by comparing them to the observed bandwidth distributions during the real operation. For this comparison, we employ the Kullback-Leibler (KL) divergence to infer the crowd size that best fits the observed distribution.

Specifically, the KL divergence between the observed bandwidth distribution $f_{obs}(x)$ and a given prior distribution $f_{BW,N}(x)$ is defined as:

\begin{equation}
    \label{eq:KL}
    D_{\mathrm{KL}}(f_{obs}(x) \parallel f_{BW,N}(x)) = \int f_{obs}(x) \log \left( \frac{f_{obs}(x)}{f_{BW,N}(x)} \right) \, dx
\end{equation}
The $N$ that minimizes the KL divergence is then selected as the best estimate of the true crowd size.

\subsection{Robustness to External Disturbances via Anomaly Detection} 
\label{anomalydetectionneed}
Although our primary objective is to estimate the number of seated individuals engaged in an activity, it is essential that the pipeline remains robust to other sources of motion in the environment. Transient events such as a person walking by, entering or exiting the room, or engaging in unrelated activities can introduce bandwidth observations that are not typical fidgeting behavior of a seated crowd. To address this, we introduce an anomaly detection module that filters out such events prior to experimental PDF construction, based on learning from synthetic bandwidth data created from our online video dataset.

More specifically, our approach relies on the intuition that motion patterns outside the scope of typical seated fidgeting will result in bandwidth sequences that the model has not seen during training. We then use a reconstruction-based anomaly detection method in which a lightweight autoencoder is trained to reconstruct expected bandwidth signals. This unsupervised learning-based approach is trained exclusively on normal (fidget-induced) data, without requiring labeled anomaly samples. Such unsupervised reconstruction-based techniques are particularly well suited for our scenario, as the anomalies are diverse, unstructured, and difficult to define categorically, which makes them impractical to label or model explicitly. Moreover, our module is trained on synthetic fidget-induced bandwidth data solely generated from available online videos, using Lemma~\ref{lem:videoBW} and a method similar to what we proposed in Section~\ref{video-to-rf}. Then, during inference, unseen patterns, such as a person walking around, yield large reconstruction errors and are flagged as anomalies to be excluded from further analysis. This module serves as both a robustness layer and a temporal denoiser, preventing non-fidget motion from distorting the crowd size estimate. To the best of our knowledge, this is the first stationary crowd-counting approach that explicitly accounts for such real-world disturbances. We note that although we are using the term "anomaly" or "disturbance", these surrounding non-fidget motions are very common for real-world seated scenarios. In Section~\ref{anomalydetection}, we will discuss in more detail how we train this module by only using video to RF-bandwidth synthetic data.

\begin{table*}[t]
\centering
\caption{Crowd counting results across 42 experiments conducted in two locations: office and conference room. }
{\footnotesize
\resizebox{\textwidth}{!}{
\begin{tabular}{B c B *{60}{c|}}
    \thickhline
    \multirow{2}{*}{\textbf{Office}} &
    \multicolumn{1}{c B}{\textbf{Real Crowd Size}} &
    \multicolumn{9}{c B}{1} &
    \multicolumn{3}{c B}{2} &
    \multicolumn{5}{c B}{3} &
    \multicolumn{2}{c B}{4} &
    \multicolumn{2}{c B}{5} \\
    \Xcline{2-23}{1.4pt}
     &
    \multicolumn{1}{c B}{\textbf{Est. Crowd Size}} &
    1 & 1 & 1 & 1 & 1 & 1 & 1 & 2 & \multicolumn{1}{c B}{2} &
    2 & 2 & \multicolumn{1}{c B}{3} &
    3 & 3 & 4 & 5 & \multicolumn{1}{c B}{7} &
    3 & \multicolumn{1}{c B}{5} &
    5 & \multicolumn{1}{c B}{7} \\
    \thickhline
    \multicolumn{23}{c}{\rule{0pt}{1pt}} \\
    \thickhline
    \multirow{2}{*}{\textbf{Conf. Room}} &
    \multicolumn{1}{c B}{\textbf{Real Crowd Size}} &
    \multicolumn{2}{c B}{1} &
    \multicolumn{3}{c B}{2} &
    \multicolumn{1}{c B}{3} &
    \multicolumn{1}{c B}{4} &
    \multicolumn{1}{c B}{5} &
    \multicolumn{2}{c B}{6} &
    \multicolumn{1}{c B}{7} &
    \multicolumn{2}{c B}{8} &
    \multicolumn{1}{c B}{9} &
    \multicolumn{2}{c B}{10} &
    \multicolumn{2}{c B}{11} &
    \multicolumn{1}{c B}{12} &
    \multicolumn{2}{c B}{13} \\
    \Xcline{2-23}{1.4pt}
    &
    \multicolumn{1}{c B}{\textbf{Est. Crowd Size}} &
    1 & \multicolumn{1}{c B}{1} &
    2 & 3 & \multicolumn{1}{c B}{3} & 
    \multicolumn{1}{c B}{3} &
    \multicolumn{1}{c B}{4} &
    \multicolumn{1}{c B}{5} &
    8 & \multicolumn{1}{c B}{8} &
    \multicolumn{1}{c B}{10} &
    11 & \multicolumn{1}{c B}{11} &
    \multicolumn{1}{c B}{9} &
    8 & \multicolumn{1}{c B}{8} &
    9 & \multicolumn{1}{c B}{10} &
    \multicolumn{1}{c B}{10} & 
    9 & \multicolumn{1}{c B}{11} \\
    \thickhline
\end{tabular}
}
}
\label{table:results}
\end{table*}

\section{Experiment Setup} \label{sec:expsetup}
In this section, we lay out our experiment setup and data collection details. We additionally discuss how the anomaly detection module is trained. 

\subsection{Experiment Settings} \label{sec:expsettings}
We conducted our experiments in two distinct indoor environments: a conference room and an office space, as shown in Fig.~\ref{fig:1}. We chose indoor spaces, as they present considerably more challenges due to the higher level of multipath, as compared to outdoor spaces. Moreover, the two indoor spaces are further chosen to represent different degrees of multipath. The conference room \rebuttal{($8$ m x $10.5$ m)} is moderately furnished and exhibits a smaller level of multipath propagation, thus offering a relatively clean RF environment. In contrast, the office \rebuttal{($3$ m x $4$ m)} is densely populated with metallic furniture, shelves, and electronics, resulting in a highly cluttered environment with significant multipath effects. This diversity allows us to evaluate the robustness of our approach in both relatively challenging and considerably challenging RF conditions.

In each environment, we conducted a series of experiments with varying crowd sizes and seating arrangements. Specifically, a subset of 13 participants,  \rebuttal{including 11 males and 2 females}, is recruited and asked to be seated for reading. The configuration of the seating, including the number of rows of chairs, spacing between individuals, and alignment with respect to the TX-RX link, is varied across trials to ensure generalization across spatial layouts. \rebuttal{The participant distance from the TX-RX link varied between 1.5 m and 4.5 m, and the minimum seat spacing between participants ranged from 0.5 m to 1 m across all experiments.} Each experiment lasted around 3 minutes. We note that the small variation around experiment times is due to experimental conditions, external interventions, and participant preparations. \rebuttal{The experiments were held across multiple days.}

\subsection{Human-Subject Protocol and Recruitment Process} \label{sec:IRB}
We note that our Institutional Review Board (IRB) committee has reviewed and approved this research. The participants were given a written form that detailed the purpose of the study, experimental procedures, data collection process, and compensation. Overall, our procedure ensured that all participants were fully informed and that the study adhered to our IRB guidelines.

\subsection{WiFi Processing Pipeline} \label{sec:wifiprocessing}
We use the internal antennas of two laptops equipped with Intel 5300 WLAN cards as transceivers. The laptops are placed 1 m above the ground and 0.5 m apart. The distance between the participants and the midpoint of the TX-RX link ranged from 1.5 m to 4.5 m during the experiments. Channel state information is collected on 30 subcarriers at 5.32 GHz, using the 802.11n standard with a transmission rate of 200 packets per second. We use one internal antenna on the transmitting laptop and all three antennas on the receiving laptop, resulting in 90 CSI data streams (30 subcarriers × 3 RX antennas). We then extract the complex CSI from these streams, using the CSI Tool \cite{csitool2011}, and generate two phase difference signals between successive receiver antennas, per subcarrier. We further denoise these 60 phase difference signals using Principal Component Analysis (PCA). More specifically, we generate the spectrograms of the first 5 PCA components, using $T_{mov}$ of 1 sec, with a shift of 0.01 sec. The five spectrograms are averaged to produce the final spectrogram. To extract the bandwidth at each time step, we compute the frequency below which 95\% of the total spectral power is concentrated.

\subsection{Autoencoder Training for Anomaly Detection} \label{anomalydetection}
The training data consists of two-second bandwidth segments derived from the video-to-bandwidth pipeline described in Section~\ref{video-to-rf}. To simulate realistic multi-person scenarios, we generate synthetic bandwidth crowd samples by randomly selecting $N$ single-person synthetic bandwidth segments (with $N$ ranging from 1 to 20), applying time-axis randomization, and aggregating the signals using a maximum operation. This process yields an empirical training set that captures the expected range and temporal dynamics of bandwidth patterns associated with the natural fidgeting of a crowd. In total, our dataset comprises approximately 600,000 such segments.

We then train a single-layer feedforward autoencoder with a hidden layer of size 32, using both L2 regularization and sparsity constraints. The network is trained for 50 epochs using the empirical bandwidth dataset. During the inference stage, the anomaly detector operates using a sliding two-second window. Each window is passed through the trained autoencoder, and the reconstruction error is computed. Any window exhibiting a reconstruction error greater than 50\% above the average training reconstruction error is flagged as anomalous and excluded from further analysis. We note that the majority of the anomalies during our experiments were caused by external people walking by. However, since our anomaly detector is based on unsupervised learning, our proposed method has the potential to exclude all types of non-fidget anomalies as part of future work. Finally, while we propose using a lightweight autoencoder-based approach, more computationally expensive machine learning architectures such as recurrent neural networks or transformer-based models could potentially enhance this module's capabilities.

\begin{table*}[t]
\centering
\caption{Crowd counting results when the anomaly detection module is disabled. The non-fidget motions, such as other people walking nearby, cause over-counting compared to Table~\ref{table:results}. Overall, the MAE of our pipeline increases from 1.04 to 1.45, showcasing that the module increases robustness in real scenarios.}
{\footnotesize
\resizebox{\textwidth}{!}{
\begin{tabular}{B c B*{60}{c|}}
    \thickhline
    \multirow{2}{*}{\textbf{Office}} &
    \multicolumn{1}{|c B}{\textbf{Real Crowd Size}} &
    \multicolumn{9}{c B}{1} &
    \multicolumn{3}{c B}{2} &
    \multicolumn{5}{c B}{3} &
    \multicolumn{2}{c B}{4} &
    \multicolumn{2}{c B}{5} \\
    \Xcline{2-23}{1.4pt}
     &
    \multicolumn{1}{c B}{\textbf{Est. Crowd Size}} &
    1 & 1 & 1 & 1 & 2 & 2 & 2 & 2 & \multicolumn{1}{c B}{2} &
    2 & 3 & \multicolumn{1}{c B}{3} &
    4 & 4 & 5 & 6 & \multicolumn{1}{c B}{7} &
    4 & \multicolumn{1}{c B}{5} &
    6 & \multicolumn{1}{c B}{7} \\
    \thickhline
    \multicolumn{23}{c}{\rule{0pt}{1pt}} \\
    \thickhline
    \multirow{2}{*}{\textbf{Conf. Room}} &
    \multicolumn{1}{|c B}{\textbf{Real Crowd Size}} &
    \multicolumn{2}{c B}{1} &
    \multicolumn{3}{c B}{2} &
    \multicolumn{1}{c B}{3} &
    \multicolumn{1}{c B}{4} &
    \multicolumn{1}{c B}{5} &
    \multicolumn{2}{c B}{6} &
    \multicolumn{1}{c B}{7} &
    \multicolumn{2}{c B}{8} &
    \multicolumn{1}{c B}{9} &
    \multicolumn{2}{c B}{10} &
    \multicolumn{2}{c B}{11} &
    \multicolumn{1}{c B}{12} &
    \multicolumn{2}{c B}{13} \\
    \Xcline{2-23}{1.4pt}
    &
    \multicolumn{1}{c B}{\textbf{Est. Crowd Size}} &
    1 & \multicolumn{1}{c B}{1} &
    3 & 3 & \multicolumn{1}{c B}{4} & 
    \multicolumn{1}{c B}{4} & 
    \multicolumn{1}{c B}{5} &
    \multicolumn{1}{c B}{6} &
    8 & \multicolumn{1}{c B}{13} &
    \multicolumn{1}{c B}{11} &
    13 & \multicolumn{1}{c B}{12} &
    \multicolumn{1}{c B}{9} &
    9 & \multicolumn{1}{c B}{10} &
    10 & \multicolumn{1}{c B}{10} &
    \multicolumn{1}{c B}{10} &
    10 & \multicolumn{1}{c B}{11} \\
    \thickhline
\end{tabular}
}
}
\label{table:results_noAE}
\end{table*}

\begin{table*}[t]
    \centering
    \caption{Comparison of different distribution distance metrics in terms of MAE, NMSE, and average convergence times.}
    \begin{tabular}{|c|c|c|c|l|}
        \hline
        Distance Measure & MAE & NMSE & Ave. Conv. Time \\
        \hline
        JS Divergence  $ \left( \frac{1}{2} D_{\mathrm{KL}}(P \,\|\, M) + \frac{1}{2} D_{\mathrm{KL}}(Q \,\|\, M), M=\frac{P+Q}{2} \right)$ & 0.90 & 0.11 & 62.00 sec  \\
        TV Distance  $ \left( \frac{1}{2} \sum |P(x) - Q(x)| \right)$ & 1.26 & 0.23 & 48.47 sec  \\
        Bhattacharyya Distance  $ \left( - \ln \left( \sum \sqrt{P(x) Q(x)} \right) \right)$ & 0.92 & 0.07 & 62.71 sec \\
        KL Divergence $ \left( \sum P(x) \log \frac{P(x)}{Q(x)} \right)$ & 1.04 & 0.15 & 50.19 sec \\
        \hline
    \end{tabular}
    \label{table:distances}
\end{table*}

\section{Results}  \label{sec:results}
In this section, we systematically evaluate the performance of our proposed pipeline through two validation methods. First, we report results from a total of 42 real-world experiments conducted across two distinct indoor environments with up to and including 13 people. Next, we present simulation results based on the video-to-bandwidth dataset to assess the performance of our proposed pipeline with even larger crowd sizes.

\subsection{Experiment Results}
In this section, we present the experimental evaluation of the proposed stationary crowd-counting pipeline.
\subsubsection{Overall Counting Performance}
\label{subsec:results}
We conducted a total of $42$ crowd reading experiments across two distinct environments: an office area and a conference room (see Fig.~\ref{fig:1}), with $21$ unique experiments performed in each location. The distribution of crowd sizes across all experiments is illustrated in Table~\ref{table:results}. We note that each experiment featured distinct seating arrangements and participant-to-transceiver distances, ensuring a diverse set of test conditions.

To evaluate counting performance, we define the error $e$ as the absolute difference between the estimated and the true number of individuals. We adopt two standard metrics for performance evaluation: the \textbf{mean absolute error (MAE)}, defined as $\mathrm{MAE} = \mathbb{E}(e)$, and the \textbf{normalized mean square error (NMSE)}, defined as $\mathrm{NMSE} = \mathbb{E}\left( \frac{e^2}{N_{\text{true}}^2} \right)$. The overall performance of our pipeline, averaged over two locations and $42$ experiments, is as follows: $\mathrm{MAE} = 1.04$ and $\mathrm{NMSE} = 0.15$. More specifically, in the office setting, the pipeline achieves $\mathrm{MAE} = 0.66$ and $\mathrm{NMSE} = 0.23$, whereas in the conference room environment, it achieves $\mathrm{MAE} = 1.42$ and $\mathrm{NMSE} = 0.06$. Table~\ref{table:results} further provides a detailed view of the estimated vs. ground-truth crowd sizes for each individual experiment. Overall, the system achieves an average miscount of approximately one person. It is worth noting that our proposed approach mainly relies on the proposed bandwidth-based solution, the associated mathematical modeling of Eq.~\ref{eq:videoBW}, and online videos of seated individuals for obtaining the bandwidth PDF of an individual.  Overall, the results highlight the robustness and generalizability of our proposed approach.

\subsubsection{Convergence Time Analysis}
A key advantage of our bandwidth-based approach over traditional fidget duty-cycle methods is that it requires less time to converge.  More specifically, the state-of-the-art requires observing fidget and silent periods for a longer period of time to accumulate reliable statistics, since each period only provides binary information. Instead, our method relies on the non-binary bandwidth distribution, resulting in a more time-efficient estimation process. We next methodically compare the convergence time with the state of the art. 

Let $\hat{N}(t)$ denote the estimated crowd size at time $t$, and let $\hat{N}_\text{final} = \hat{N}(T)$ be the final estimate at the end of the observation window $T$. We define the \textbf{convergence time} as $t_\text{conv} = \max \left\{ t \in [0, T) \,\middle|\, \left| \hat{N}(t) - \hat{N}_\text{final} \right| > 1 \right\}.$ This definition effectively captures the time it takes for a pipeline to converge to a stable estimate for a given experiment.

\textbf{Comparison with state-of-the-art:} 
Our pipeline converges to a stable crowd size estimate within an average time of $51$ seconds, when averaged over all the $42$ experiments. This indicates a substantial improvement over state-of-the-art fidget duty cycle-based method. For instance, \cite{korany2021counting} reports estimation vs. time plots for four of their experiments (with varying crowd sizes from 4 to 10). We can estimate an average convergence time of $112.5$ seconds from these plots, which is considerably higher than our average convergence time of $51$ seconds. As discussed earlier, this is due to the fact that our pipeline utilizes the bandwidth of the signal, which enables the accumulation of the needed statistical information faster.

\subsubsection{Effect of Anomaly Detection Module}
We next evaluate the impact of the anomaly detection module introduced in Section~\ref{anomalydetectionneed}. Table~\ref{table:results_noAE} shows the estimated vs. ground truth crowd sizes across all experiments when the anomaly detection module is disabled. The absence of anomaly filtering leads to noticeable over-counting, primarily due to transient disturbances and non-fidget motions such as people walking outside the experiment room. Quantitatively, the system without anomaly detection achieves $\mathrm{MAE} = 1.45$ and $\mathrm{NMSE} = 0.32$ (as compared to $\mathrm{MAE} = 1.04$  and $\mathrm{NMSE} = 0.15$ of Section~\ref{subsec:results} for when we deploy anomaly detection). These results indicate a degradation in accuracy, emphasizing the importance of the anomaly detection module.

\subsubsection{Effect of Distribution Distance Measure Selection}
Next, we evaluate the impact of the chosen distance metric for comparing the observed and prior bandwidth distributions. Specifically, we consider four commonly used probability distribution distance measures: Kullback–Leibler (KL) divergence, Jensen–Shannon (JS) divergence, total variation (TV) distance, and Bhattacharyya distance. Table~\ref{table:distances} summarizes the performance of each metric in terms of MAE, NMSE, and average convergence time. Our pipeline demonstrates consistent performance across all distance metrics, indicating robustness to the choice of divergence measure.  As can be seen, there is a trade-off between error and average convergence times. For instance, the Bhattacharyya distances achieve the lowest MAE of $0.92$ and an NMSE of $0.07$  while having the highest average convergence time of $62.71$ seconds. In contrast, the TV distance achieves the lowest average convergence time of $48.47$ seconds while having the highest MAE and NMSE of $1.26$ and $0.23$, respectively. Overall, the selected KL divergence metric strikes a good balance between accuracy and convergence speed.

\begin{figure*}[t]
    \centering
    \includegraphics[width=\textwidth]{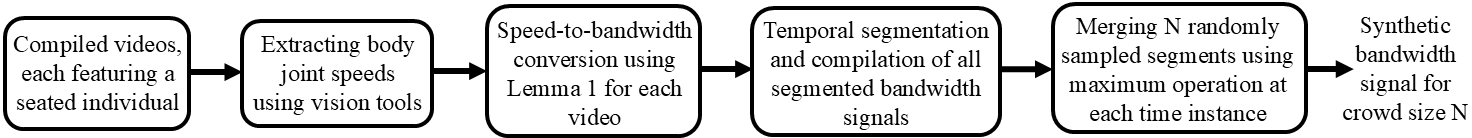} 
    \caption{Proposed pipeline to synthetically generate bandwidth data for different crowd sizes using readily-available videos, each of a seated individual, and our video-to-bandwidth approach. This pipeline complements our extensive experimentation of 42 tests with up to and including 13 individuals, and enables testing with larger crowds in an efficient simulation environment.}
    \label{fig:simulationpipeline}
\end{figure*}

\begin{table*}[t]
\centering
\caption{Crowd counting results for empirical simulation dataset. For each crowd size, the MAE for 296 test samples is reported.}
{\footnotesize
\resizebox{\textwidth}{!}{
\begin{tabular}{|c|*{60}{c|}}
    \hline
    \multicolumn{1}{|c|}{\textbf{Crowd Size}} &
    \multicolumn{1}{c|}{1} &
    \multicolumn{1}{c|}{2} &
    \multicolumn{1}{c|}{3} &
    \multicolumn{1}{c|}{4} &
    \multicolumn{1}{c|}{5} &
    \multicolumn{1}{c|}{6} &
    \multicolumn{1}{c|}{7} &
    \multicolumn{1}{c|}{8} &
    \multicolumn{1}{c|}{9} &
    \multicolumn{1}{c|}{10} &
    \multicolumn{1}{c|}{11} &
    \multicolumn{1}{c|}{12} &
    \multicolumn{1}{c|}{13} &
    \multicolumn{1}{c|}{14} &
    \multicolumn{1}{c|}{15} &
    \multicolumn{1}{c|}{16} &
    \multicolumn{1}{c|}{17} &
    \multicolumn{1}{c|}{18} &
    \multicolumn{1}{c|}{19} &
    \multicolumn{1}{c|}{20} \\
    \hline   
    \textbf{MAE} & 
    0.24 & 0.30 & 0.46 & 0.61 & 0.70 & 0.84 & 0.96 & 1.15 & 1.21 & 1.34 &  1.46 & 1.60 & 1.78 & 1.91 & 2.10 & 2.23 & 2.38 & 2.56 & 2.72 & 2.89 \\
    \hline
\end{tabular}
}
}
\label{table:simulationresults}
\end{table*}

\subsection{Simulation-Based Results to Test Larger Crowd Sizes}
So far, we have shown real experimental results with crowds up to and including 13 individuals. \rebuttal{To showcase our pipeline's performance with even larger crowds, we next present a novel framework to efficiently simulate and test with larger crowd sizes. \textbf{This pipeline synthesizes crowd bandwidths directly from readily available videos. Leveraging this video-to-bandwidth modeling eliminates the need for a full-body crowd wave simulator} (such as those used in \cite{cai2020teaching}), which is difficult to make fully realistic in terms of wave interactions.}  

More specifically, to simulate synthetic bandwidth observations for
larger crowd sizes, we use the video-to-bandwidth pipeline of
Section IV-C and public videos of single seated individuals.
Each video is segmented into 3-minute non-overlapping segments. Each segment is then processed through the video-tobandwidth pipeline, which yields the corresponding bandwidth measurements for the seated individual as if a transceiver
were present. To synthetically generate test data for any crowd size $N$, we then randomly sample $N$ of these segments and combine them by taking the maximum bandwidth value across all $N$ segments at each time instance. This approach efficiently models the observed bandwidth expected from a group of $N$ independently fidgeting, seated individuals, where the overall observed bandwidth is dictated by the highest contributor at any given time. \rebuttal{This aggregation mimics the dominant bandwidth behavior of real multi-person environments.} We repeat this sampling strategy for $N \in {2,3, ..., 20}$ with as many synthetic test samples generated per $N$ as the number of unique combinations that can be created from the video pool. Fig.~\ref{fig:simulationpipeline} summarizes our proposed method for synthetic test data generation. 

To ensure a strict separation between prior individual PDF generation of Section~\ref{sec:systemdesign}, which is needed as part of the mathematical modeling, and seated crowd data for testing in a simulation environment, we adopt 3-fold cross-validation over our full video dataset of 32 unique videos of individuals. In each fold, approximately 21 videos are used for prior generation purposes, whereas the remaining 11 are used for testing in a simulation environment. The prior pool is then used to construct prior individual PDF distributions following the proposed method in Section~\ref{sec:systemdesign}, while the test pool is used to generate the synthetic bandwidth observations as described above, for testing with larger crowds in a simulation environment.

In order to mitigate the bias towards any specific random selection, we repeat the entire cross-validation process five times with different randomizations and report the average results. In total, for each crowd size, we evaluate our pipeline on $296$ unique 3-minute segments per crowd size, yielding a total test set size of $20 \times 296 = 5920$ empirical test instances. We note that each data segment is unique. However, during the extrapolation to larger crowd sizes, different segments belonging to the same video may be randomly picked as part of the same synthetic crowd data sample. 

Table~\ref{table:simulationresults} reports the MAE for each crowd size. Our system achieves an overall NMSE of $0.054$ and an overall MAE of $1.47$ over the simulated dataset. These results further showcase the potential of our bandwidth-based approach with even larger crowd sizes. 

\subsection{Computation Time}
On an Intel Core i7-9700K processor, our pipeline takes $69$ ms, on average, to process one second of data.

\section{Discussion and Future Directions} \label{discussion}
\rebuttal{
We next discuss several aspects of our proposed work and possible future directions.
}
\subsection{Impact of $\psi$}
\rebuttal{As mentioned in Sec.~\ref{sec:wifiprocessing}, the TX-RX separation is intentionally chosen to be small (0.5 m) in our experiments. This design choice ensures that $\psi$ remains very close to its upper bound (2) for all realistic crowd distances. For example, when a person is as close as 1.5 m to the link, $\psi = 1.97$. Such a close TX-RX setup and the corresponding $\psi \approx 2$ is commonly used in the RF sensing literature. The high performance of our pipeline further confirms the validity of this assumption in our setup. If the TX and RX are not close to each other, $\psi$ can deviate from 2. RF sensing under such a setup is less studied and constitutes an important avenue for future work. }

\subsection{Independence of the received bandwidth across individuals}
\rebuttal{
In Sec.~\ref{bw-to-crowd}, we took the fidgeting patterns of different individuals in a seated crowd to be uncorrelated. To validate this assumption, we conducted two experiments in which six participants were video recorded while reading. Using the MediaPipe pose extractor together with the proposed video-to-bandwidth pipeline (Sec.~\ref{sec:systemdesign}), we generated individual bandwidth time series and computed pairwise Pearson correlation coefficients across participants. The average correlation coefficient magnitude was $0.142$ in the first experiment and $0.118$ in the second one. The resulting low correlation values further confirm that individual fidgeting behaviors are largely uncorrelated across participants, supporting the i.i.d.~assumption of Sec.~\ref{bw-to-crowd}.
}

\subsection{Impact of the anomaly detector threshold}
\begin{figure}[t]
    \centering
    \includegraphics[width=0.48\textwidth]{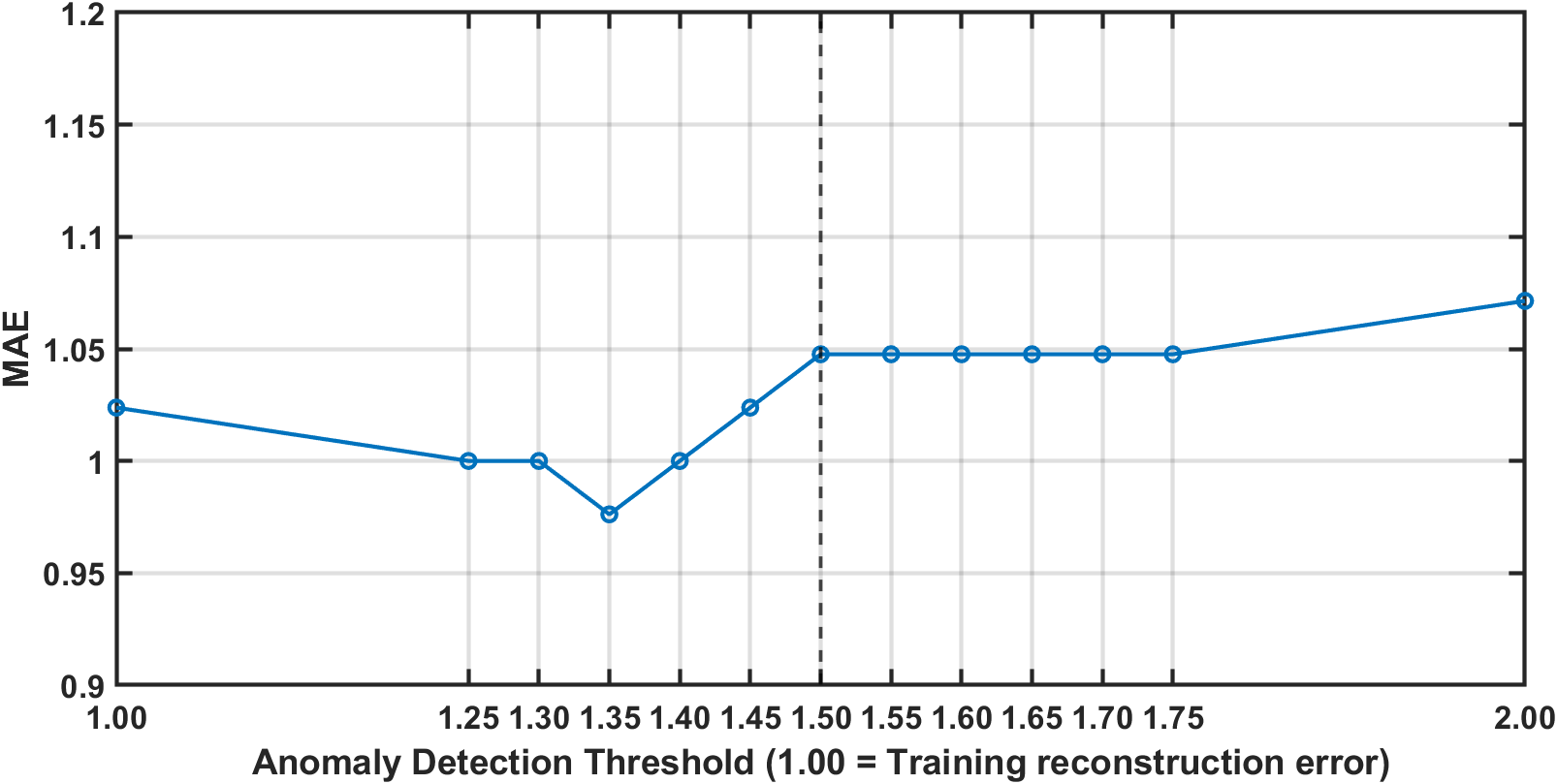} 
    \caption{\rebuttal{Sensitivity of the proposed pipeline to the anomaly detection threshold. The threshold is expressed as a multiple of the average training reconstruction error (e.g., 1.50 corresponds to 50\% above the training reconstruction error). MAE remains stable over a wide range of thresholds, demonstrating the robustness of the proposed pipeline.}}
    \label{fig:thresholdsensitivity}
\end{figure}

\rebuttal{
In Sec.~\ref{anomalydetection}, we set the anomaly detection rule such that any window exhibiting a reconstruction error greater than 50\% above the average training reconstruction error is flagged as anomalous and excluded from further analysis. In this section, we conduct a sensitivity analysis by varying the threshold across a range of values. Figure \ref{fig:thresholdsensitivity} shows the resulting MAE of our pipeline as the anomaly detection threshold is swept from 1.0 to 2.0 times the training reconstruction error. As can be seen, our pipeline's performance is robust with respect to the threshold, with the MAE remaining within the narrow range of 0.97–1.04 when the anomaly detection threshold lies between 1.25 and 1.75 times the training reconstruction error.
}

\subsection{Applicability Beyond Reading}
\rebuttal{
While our experiments focus on cases where the crowd is engaged in reading, the proposed pipeline does not rely on reading as the activity. Instead, we have proposed a method to exploit the fidget statistics of a seated crowd to infer its size. Therefore, the proposed framework can generalize to other seated activities (e.g., studying, typing, or watching a presentation) by utilizing the relevant activity videos.
}

\section{Conclusions}  \label{sec:conclusion}
In this paper, we considered the challenging problem of counting stationary crowds with commodity WiFi transceivers. We proposed a new approach that links the bandwidth of the received signal near a seated crowd to the crowd size by leveraging the aggregate natural fidgeting behavior of individuals. To characterize the individual fidgeting PDF required for our mathematical model, we proposed a novel method that first extracts body speed profiles from publicly available videos of a seated individual using vision techniques. We then apply a speed-to-bandwidth conversion, inspired by Carson’s Rule from analog FM radio design, to obtain the corresponding fidgeting bandwidth PDF of an individual. Finally, we introduced an anomaly detection module in order to capture and exclude any non-fidget motions, such as a bystander walking by. We conducted $42$ WiFi experiments with crowd sizes up to and including 13 individuals, and in two indoor environments. Our results show that the proposed approach can estimate crowd sizes well with an MAE of 1.04 and an NMSE of 0.15, and an average convergence time of only 51 seconds, significantly reducing the convergence time as compared to the state of the art. Overall, the paper enables fast, robust, and highly accurate counting of seated crowds, with commodity WiFi signals.

\section*{References}
\bibliography{main}
\bibliographystyle{IEEEtran}

\end{document}